\documentclass{jfm}

\usepackage{graphicx}
\usepackage{newtxtext}
\usepackage{newtxmath}
\usepackage{natbib}
\usepackage{hyperref}
\hypersetup{
    colorlinks = true,
    urlcolor   = blue,
    citecolor  = blue,
}

\newcommand{\RomanNumeralCaps}[1]
\linenumbers

\title{Numerical study on the mechanism of drag modulation by dispersed drops in two-phase Taylor-Couette turbulence}

\author{Jinghong Su\aff{1}, Lei Yi\aff{1}, Bidan Zhao\aff{2}, Cheng Wang\aff{1}, Fan Xu\aff{2}, Junwu Wang\aff{2}\corresp{\email{jwwang@ipe.ac.cn}}, \and Chao Sun\aff{1,3}   \corresp{\email{chaosun@tsinghua.edu.cn}}}

\affiliation{\aff{1}New Cornerstone Science Laboratory, Center for Combustion Energy, Key Laboratory for Thermal Science and Power Engineering of Ministry of Education, Department of Energy and Power Engineering, Tsinghua University, 100084 Beijing, China
\aff{2}State Key Laboratory of Multiphase Complex Systems, Institute of Process Engineering, Chinese Academy of Sciences, P. O. Box 353, Beijing 100190, PR China
\aff{3}Department of Engineering Mechanics, School of Aerospace Engineering, Tsinghua University, Beijing 100084, China}

\begin{document}
\maketitle

\begin{abstract}
The presence of a dispersed phase can significantly modulate the drag in turbulent systems. We derived a conserved quantity that characterizes the radial transport of azimuthal momentum in the fluid-fluid two-phase Taylor-Couette turbulence. This quantity consists of contributions from advection, diffusion, and two-phase interface, which are closely related to density, viscosity, and interfacial tension, respectively.
We found from interface-resolved direct numerical simulations that the presence of the two-phase interface consistently produces a positive contribution to the momentum transport and leads to drag enhancement, while decreasing the density and viscosity ratios of the dispersed phase to the continuous phase reduces the contribution of local advection and diffusion terms to the momentum transport, respectively, resulting in drag reduction. 
Therefore, we concluded that the decreased density ratio and the decreased viscosity ratio work together to compete with the presence of two-phase interface for achieving drag modulation in fluid-fluid two-phase turbulence.
\end{abstract}

\begin{keywords}
%Authors should not enter keywords on the manuscript, as these must be chosen by the author during the online submission process and will then be added during the typesetting process (see \href{https://www.cambridge.org/core/journals/journal-of-fluid-mechanics/information/list-of-keywords}{Keyword PDF} for the full list).  Other classifications will be added at the same time
\end{keywords}

% {\bf MSC Codes }  {\it(Optional)} Please enter your MSC Codes here

\section{Introduction}
\label{sec:intro}
Two-phase flow, consisting of two immiscible fluids, is widely encountered in various engineering applications. The presence of a dispersed phase can significantly alter the flow characteristics, leading to either drag enhancement or drag reduction~\citep{ceccio2010friction,balachandar2010turbulent,lohse2018bubble, mathai2020bubbly, yi2023recent}. 
A correct understanding of the mechanism of drag modulation is of great significance to relevant engineering applications; however, a comprehensive understanding of this mechanism is still missing.

A typical characteristic of liquid-liquid two-phase flow is the dynamics of the interface, which involves deformation~\citep{rallison1984deformation, rosti2019numerical, haakansson2022criterion}, coalescence~\citep{stone1994dynamics, kavehpour2015coalescence}, and breakup~\citep{lemenand2017turbulent, olad2023comparison,ni2023deformation}. When coalescence is counterbalanced by breakup, the dispersed phase exhibits a specific size distribution which can be well described by a lognormal distribution~\citep{yi2021global}. Under fixed Reynolds number, the dispersed phase has almost the same average size for different volume fractions~\citep{yi2022physical}.
It has been observed that the dispersed phase volume fraction is positively correlated with the global transport of a liquid-liquid two-phase turbulent flow~\citep{yi2022physical}, indicating that increasing the interfacial area could contribute to the drag enhancement.  
The effect of dispersed phase coalescence in liquid-liquid two-phase flow has recently been studied using interface-resolved direct numerical simulations~\citep{de2019effect,cannon2021effect}, and it is found that the coalescence effectively decreases the interfacial area, thus weakening the drag enhancement effect, and vice versa~\citep{de2019effect}. Additionally, the interface deformation is also found to be important for the gas-liquid turbulent drag reduction~\citep{van2005drag, van2013importance,spandan2017deformable,lohse2018bubble,mathai2020bubbly}. The bubbles with large size or high deformability exhibit pronounced drag reduction effect~\citep{lu2005effect,verschoof2016bubble}, but the drag reduction effect is lost when the bubbles are reduced to small sizes after the addition of surfactants~\citep{verschoof2016bubble}.

Existing studies show that interface dynamics induce very different drag modulation effects in liquid-liquid and gas-liquid two-phase flows ~\citep{lohse2018bubble,mathai2020bubbly,yi2023recent}. However, the role played by the fluid properties of the dispersed phase in these effects remains unclear. 
The interface dynamics, density, and viscosity of the dispersed phase are intricately coupled, making it challenging to isolate the influence of the individual effects on drag modulation.
In this study, we utilize interface-resolved three-dimensional direct numerical simulations to track interface dynamics in Taylor-Couette turbulence. Additionally, we employ momentum budget analysis to investigate the individual and coupling effects of the interface dynamics, density, and viscosity of the dispersed phase. We aim to uncover how these factors operate and determine whether they cooperate or compete with each other in drag modulation.

\section {Results and discussion}
 The interface-resolved three-dimensional direct numerical simulations of two-phase fluid-fluid flow in a Taylor-Couette (TC) system were carried out using a volume-of-fluid (VOF) method with the piecewise linear interface calculation (PLIC) based on the open-source OpenFOAM v8~\citep{rusche2003computational,chen2022turbulent}.
 We consider two immiscible and incompressible fluids confined between two coaxial cylinders whose radii are $r_i$ (inner) and $r_o$ (outer). In this work, we have chosen to fix the outer cylinder while allowing the inner cylinder to rotate with a constant angular velocity $\omega_i$.
The two-phase flow is governed by the Navier-Stokes equations 
\begin{equation}
  {\partial _t \rho}
  +\nabla\cdot(\rho \boldsymbol{u})
  =0,
  \vspace{-0mm}
\end{equation}
\begin{equation}
  {\partial _t(\rho \boldsymbol u)}
  +\nabla\cdot(\rho\boldsymbol{uu})
  =
  -{\nabla{p}}
  +\nabla\cdot{[\mu(\nabla\boldsymbol{u}+\nabla{^T\boldsymbol u})]}
  +\boldsymbol{f},
\end{equation}
where $\boldsymbol{u}$ is the velocity and $p$ is the pressure. The $\rho$ and $\mu$ are the variable density and viscosity, respectively. The continuous carrier phase is characterized by the density $\rho_f$ and viscosity $\mu_f$, whereas the dispersed phase is described by the density $\rho_d$ and viscosity $\mu_d$. The phase fraction $\alpha$ is introduced to characterize the variable density and viscosity, i.e., $\rho=\alpha\rho_d+(1-\alpha)\rho_f $ and $\mu=\alpha\mu_d+(1-\alpha)\mu_f $. 
The continuum surface force method, as proposed by \cite{brackbill1992continuum}, is adopted in this study to describe the interfacial tension, i.e., $\boldsymbol{f}=-\sigma\kappa\nabla\alpha$, where $\sigma$ denotes the surface tension coefficient and $\kappa=\nabla\cdot(\nabla\alpha/|\nabla\alpha|)$ represents the interface curvature.
\textcolor{black}{The accuracy of the VOF method is often affected by interface sharpness~\citep{gamet2020validation} and spurious currents~\citep{vachaparambil2019comparison,fan2020varrhoturbvof}, the detailed numerical methods and computational accuracy are presented in the Supplemental Material.}

\textcolor{black}{In the simulated Taylor-Couette system, a rotational symmetry of six (i.e., the simulated azimuthal region is set as $\pi/3$) is chosen to reduce computational costs without compromising the accuracy of our results. This choice has been validated for both single-phase and multiphase Taylor-Couette turbulence~\citep{brauckmann2013direct, spandan2018physical, assen2022strong, xu2022direct, xu2023direct}.} The curvature of Taylor-Couette system is defined as $\eta=r_i/r_o=0.714$ and the aspect ratio is defined as $\Gamma=L/d=2\pi/3$, where $d = r_o-r_i$ is the gap width and $L$ is the axial length. No-slip and impermeable boundary conditions are imposed in the radial direction, while periodicity is imposed in the axial and azimuthal directions. 
\textcolor{black}{The inner and outer cylinders are subjected to a Neumann boundary condition for the phase fraction, resulting in a default contact angle of $90^\circ$.}
The system is uniformly discretised by (336$\times$192$\times$192) grid points in the azimuthal, radial, and axial directions, respectively. 
\textcolor{black}{The grid spacing is measured in wall units $y^+$ based on the shear stress at the inner cylinder for single-phase flow. In the radial direction, the uniform grid spacing is $0.725y^+$. Alternatively, a total amount of 6 grids is embedded inside the viscous sublayer, whose thickness is $0.0359d$. The uniform grid spacing in the axial direction is $1.519y^+$. The grid spacing in the azimuthal direction ranges from $1.085y^+$ near the inner wall to $1.519y^+$ near the outer wall.}
The Reynolds number ${\rm Re}={\rho_f}r_i \omega_i d/\mu_f$ and the Weber number $We={\rho_f}r_i^2 \omega_i^2 d/\sigma$ are fixed at 2000 and 1260, respectively. \textcolor{black}{The Taylor number is fixed as ${Ta}=[\rm Re(1+\eta)^3/(8\eta^2)]^2=6.1\times 10^6$.} \textcolor{black}{Due to the increased computational resources required for resolving the two-phase interface and the smaller time step needed to capture capillary wave propagation~\citep{brackbill1992continuum} in the two-phase flow compared to the single-phase case, we have chosen $Ta=6.1\times 10^6$ to ensure compatibility with our available computational resources. When the Taylor number exceeds $3\times10^6$, the flow field transitions into the classical regime of Taylor-Couette turbulence~\citep{grossmann2016high}. Our chosen Taylor number falls within this regime, ensuring an adequate amount of turbulence information. The Weber number of 1260 is chosen to ensure that the spurious velocities remain within an acceptable range, as depicted in Fig. S2 in the Supplemental Material.} The maximum Courant-Friedrichs-Lewy number is set to be 0.2. 
All the presented statistics are collected for at least 20 turns over time after reaching a statistically steady state.

\begin{figure}%[htbp]
	\centering
	\includegraphics[width=0.6\linewidth]{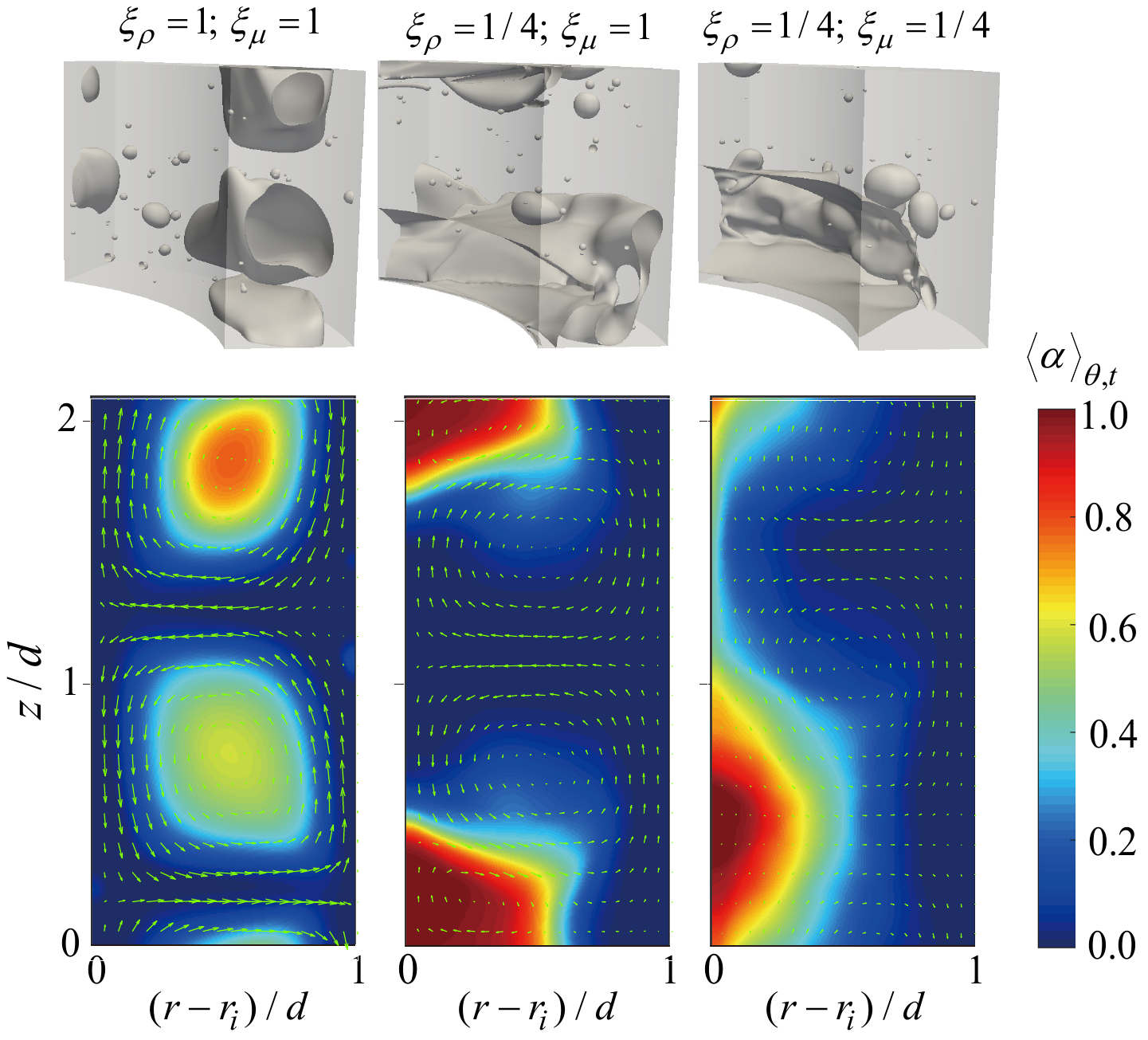}
	\caption{Instantaneous interface snapshots and corresponding azimuthally- and time-averaged phase fraction $\left \langle \alpha \right \rangle _{\theta,t}$. \textcolor{black}{The green arrows indicating the direction and magnitude of the averaged radial–axial velocity vectors $\left \langle u_{rz} \right \rangle _{\theta,t}/u_i$, where $u_i=\omega_i r_i$ is the velocity of the inner cylinder.}
	}
	\label{fig:figure1}
\end{figure}

Due to the centrifugal effect in the Taylor-Couette system, the distribution of dispersed phase is affected by the density ratio $\xi_{\rho} = \rho_d/\rho_f$ as well as the viscosity ratio  $\xi_{\mu} = \mu_d/\mu_f$ of the dispersed phase to the continuous phase. Figure~\ref{fig:figure1} displays the instantaneous interface snapshots and phase distribution for the dispersed phase volume fraction of $\varphi=20\%$, providing valuable insights into the spatial distribution and behavior of the interface and phase components in the system.  \textcolor{black}{The averaged radial–axial velocity vectors are also displayed by green arrows to denote the structure and strength of the Taylor-vortex.
When $\xi_{\rho} = 1$ and $\xi_{\mu} = 1$, the dispersed phase predominantly accumulates at the center of the Taylor vortex, which is primarily attributed to the linear shear gradient present in the system~\citep{hori2023interfacial}.
When $\xi_{\rho} = 1/4$ and $\xi_{\mu} = 1$, the dispersed phase migrates towards the inner wall of the system due to the centrifugal force exerted by the rotating flow. Consequently, the dispersed phase gathers in the plume ejection region near the inner wall and weakens the Taylor vortex.
Furthermore, when $\xi_{\rho} = 1/4$ and $\xi_{\mu} = 1/4$, the distribution of the dispersed phase becomes more concentrated near the inner wall and the Taylor vortex is again weakened. The results indicate that decreasing the density ratio $\xi_{\rho}$ and the viscosity ratio $\xi_{\mu}$ facilitates the gathering of the dispersed phase near the inner wall and the weakening of the Taylor vortex.
Our objective is to analyze the specific effects of dispersed phase density and viscosity on the process of drag reduction.}

Through sequential variations of the volume fraction $\varphi$, the density ratio $\xi_{\rho}$, and the viscosity ratio $\xi_{\mu}$, we conducted a comprehensive study to investigate the influence of these parameters on the drag modulation (see Table 1).  
The results reveal several important findings. Firstly, when $\xi_{\rho} = 1$ and $\xi_{\mu} = 1$, it causes a minor increase in drag, indicating a slight drag enhancement effect. However, when we lower the $\xi_{\rho}$ to 1/4, a significant drag reduction is observed. Moreover, further reducing $\xi_{\mu}$ to 1/4 enhances the drag reduction effect even more. Furthermore, we found that the drag modulation shows similar trends for the dispersed phase volume fraction of $\varphi=10\%$ and $\varphi=20\%$.  Under fixed density ratio $\xi_{\rho}$ and viscosity ratio $\xi_{\mu}$, an increase in the volume fraction $\varphi$ leads to a stronger drag enhancement or drag reduction effect. In other words, as the volume fraction of the dispersed phase increases, the influence on the drag becomes more pronounced. These results highlight the significant impact of the dispersed phase's density, viscosity, and volume fraction on the flow dynamics, particularly in terms of drag enhancement and drag reduction effects.

\begin{table}%[htbp]%[b]%The best place to locate the table environment is directly after its first reference in text
    	\centering
     \setlength\tabcolsep{20pt}
		\begin{tabular}{cccc}
               % \hline
			\textrm{$\varphi$}&
			\textrm{$\xi_{\rho}=\rho_d /\rho_f$}&
			\textrm{$\xi_{\mu}= \mu_d /\mu_f$}&
			\multicolumn{1}{c}{\textcolor{black}{Drag modulation ($T/T_{\varphi=0}-1$)}}\\[6pt]
			%\colrule
			%\hline
			$0  $  & ---		  & ---	 		 & ---		              \\[2pt]
			$10\%$ & $1$	  & $1$	 	& $+ 0.63\%$ 	      \\[2pt]
			$10\%$ & $1/4$ & $1$	 	& $-16.02\%$	       \\[2pt]
			$10\%$ & $1/4$ & $1/4$	 	& $-23.44\%$ 	       \\[2pt]
			$20\%$ & $1$	  & $1$	 	& $+ 1.04\%$  	       \\[2pt]
			$20\%$ & $1/4$ & $1$	 	& $-28.31\%$ 	       \\[2pt]
			$20\%$ & $1/4$ & $1/4$	 	& $-50.23\%$ 	       \\[2pt]

     % \hline
		\end{tabular}
     	\caption{\label{tab:table1}%
		Drag modulation in two-phase flow. \textcolor{black}{The $T$ represents the torque exerted on the inner cylinder and $T_{\varphi=0}$ denotes the torque specifically associated with the single-phase flow condition.}
	}
\end{table}

\begin{figure}%[htbp]
	\centering
	\includegraphics[width=0.9\linewidth]{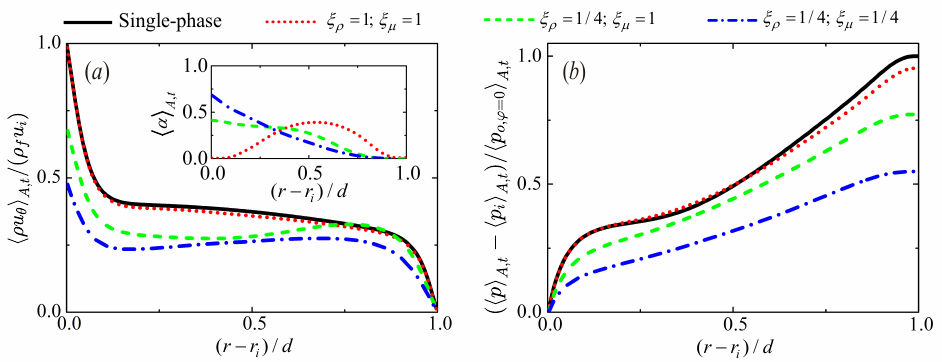}
	\caption{
		\textcolor{black}{(\textit{a}) The azimuthal momentum $\left \langle \rho u_\theta \right \rangle _{A, t}/(\rho_f u_i)$ and (\textit{b}) pressure $(\left \langle p \right \rangle_{A, t}-\left \langle p_i \right \rangle_{A, t})/\left \langle p_{o,\varphi=0} \right \rangle_{A, t}$ as a function of the radial position, where $u_\theta$ is the azimuthal velocity, $p_i$ is the pressure at the inner cylinder, and $p_{o,\varphi=0}$ is the pressure at the outer cylinder in single-phase flow. Correspondingly, the inset figure exhibits the radial profile of phase fraction $\left \langle \alpha \right \rangle _{A, t}$.  The average operator $\left \langle \cdot \right \rangle _{A, t}$ is to obtain the axially-, azimuthally- and time-averaged value of the quantity.}}
	\label{fig:azimuthal velocity}
\end{figure}

Given the similar trends in drag modulation for the dispersed phase volume fraction of $\varphi=10\%$ and $\varphi=20\%$, we next focus on the case with $\varphi=20\%$.
To investigate the impact of the dispersed phase on the flow field, 
Fig.~\ref{fig:azimuthal velocity} displays the radial profiles of the normalized azimuthal momentum and pressure calculated based on the phase distribution (see the inset figure) as a reference.
When $\xi_{\rho} = 1$ and $\xi_{\mu} = 1$, the azimuthal momentum in the bulk region of the two-phase flow shows a slight decrease. The pressure is modulated in the region away from the inner cylinder and shows a decrease near the outer cylinder. Specifically, the modulation starts at a point where the phase fraction deviates from zero, but the modulation is insignificant overall.
By reducing the density ratio $\xi_{\rho}$ or viscosity ratio $\xi_{\rho}$, a notable decrease in the azimuthal momentum and pressure in the bulk region is observed, aligning with the observed drag modulation.
It is revealed that the reduction in the density ratio and the viscosity ratio plays a vital role in causing this decrease. The density ratio and the viscosity ratio are identified as key factors influencing drag reduction. 

\begin{figure}%[htbp]
	\centering
	\includegraphics[width=0.9\linewidth]{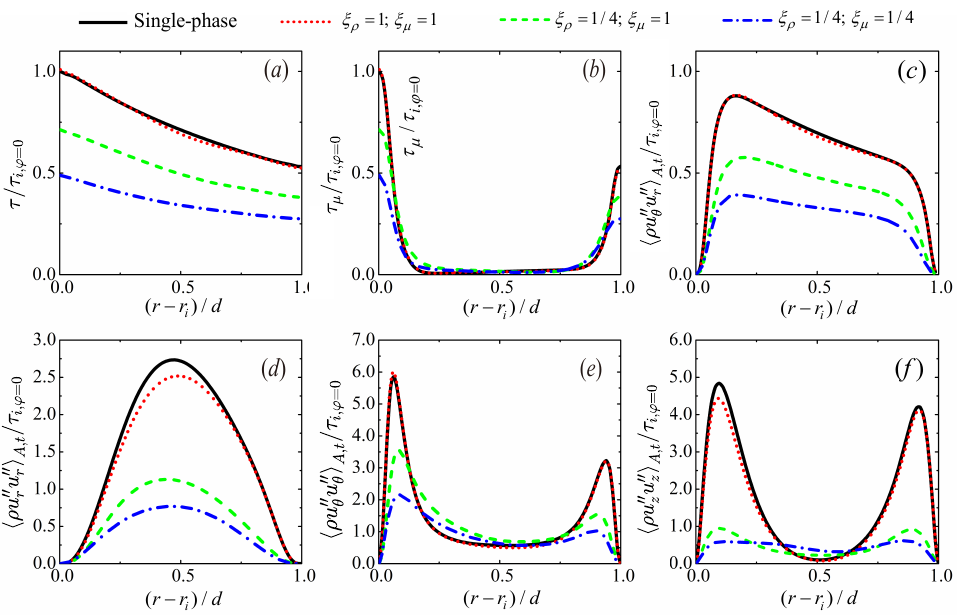}
	\caption{
		\textcolor{black}{(\textit{a}) The total shear stress 
  $\tau/\tau_{i,\varphi=0}$, (\textit{b}) viscous shear stress $\tau_\mu/\tau_{i,\varphi=0}$, and (\textit{c}) Favre shear stress $-{\left \langle {\rho u_\theta^{\prime\prime} u_r^{\prime\prime}} \right \rangle _{A, t}}/\tau_{i,\varphi=0}$ as a function of the radial position, where ${\tau}_{i,\varphi=0}$ is the total shear stress on the inner cylinder in single-phase flow. The Favre normal stresses (\textit{d}) ${\left \langle {\rho u_r^{\prime\prime} u_r^{\prime\prime}} \right \rangle _{A,t}}/\tau_{i,\varphi=0}$, (\textit{e}) ${\left \langle {\rho u_\theta^{\prime\prime} u_\theta^{\prime\prime}} \right \rangle _{A,t}}/\tau_{i,\varphi=0}$, and (\textit{f}) ${\left \langle {\rho u_z^{\prime\prime} u_z^{\prime\prime}} \right \rangle _{A,t}}/\tau_{i,\varphi=0}$ are also shown as a function of the radial position.
  }
	}
	\label{fig:stress}
\end{figure}

\textcolor{black}{To investigate the effects of dispersed phase density and viscosity on the turbulent transport, we investigate the radial profile of total shear stress (Fig.~\ref{fig:stress}a), viscous shear stress (Fig.~\ref{fig:stress}b), and their difference (Fig.~\ref{fig:stress}c). Due to the density difference between the dispersed and continuous phases, the Reynolds averaging is no longer applicable. We therefore adopt the Favre averaging~\citep{favre1969statistical}, which applies a density-weighted average to the velocity. The difference between the total shear stress and viscous shear stress is thus written as
$\tau-\tau_\mu=-{\left \langle {\rho u_\theta^{\prime\prime} u_r^{\prime\prime}} \right \rangle _{A, t}}$ (hereafter referred to as Favre shear stress), which could be used to characterize the turbulence term.
Here, 
$u_\theta^{\prime\prime}
=
u_\theta
-
\left \langle {\rho u_\theta} \right \rangle _{A,t}
/
\left \langle {\rho} \right \rangle _{A,t}$ 
and $u_r^{\prime\prime}
=
u_r
-
\left \langle {\rho u_r} \right \rangle _{A,t}
/
\left \langle {\rho} \right \rangle _{A,t}$.
For the two-phase flow with $\xi_{\rho} = 1$ and $\xi_{\mu} = 1$, the profiles of the shear stresses are nearly identical to those for the single-phase case.
%the total shear stress in the bulk region exhibits a slight decrease, aligning with the corresponding angular momentum modulation. 
However, by reducing the density ratio $\xi_{\rho}$ or viscosity ratio $\xi_{\mu}$, a significant decrease in the total shear stress is observed due to the decrease near the wall for the viscous shear stress and the decrease in the bulk region for the Favre shear stress, which is consistent with the observed drag reduction.}

\textcolor{black}{To further study the influence of the dispersed phase on the turbulence fluctuations, we show the Favre normal stresses (Figs.~\ref{fig:stress}d-\ref{fig:stress}f), which are generally investigated by the root mean square (r.m.s.) velocity fluctuations in the constant density flow system ~\citep{zhu2016direct,wang2023maximum}.
For the two-phase flow with $\xi_{\rho} = 1$ and $\xi_{\mu} = 1$, the normal stress modulation is not very significant and the profiles of the normal stresses almost coincide with those for the single-phase case. 
By reducing the density ratio $\xi_{\rho}$ or viscosity ratio $\xi_{\mu}$, there is an overall reduction in the normal stress ${\left \langle {\rho u_r^{\prime\prime} u_r^{\prime\prime}} \right \rangle _{A,t}}$. The normal stress ${\left \langle {\rho u_\theta^{\prime\prime} u_\theta^{\prime\prime}} \right \rangle _{A,t}}$ shows a significant decrease near the inner and outer cylinders, while an increase is seen in the bulk region. A similar phenomenon has also been found in the study of the r.m.s. streamwise velocity fluctuations for polymer drag reduction systems, i.e., the r.m.s. streamwise velocity fluctuations decrease near the wall but increase away from the wall~\citep{min2003drag,wang2023maximum}. 
\textcolor{black}{The normal stress ${\left \langle {\rho u_z^{\prime\prime} u_z^{\prime\prime}} \right \rangle _{A,t}}$ shows a stronger reduction, indicating that the Taylor-vortex has been significantly weakened as indicated by the averaged radial–axial velocity vectors in Fig. 1.}
As a result, the drag reduction caused by decreasing the density or viscosity of the dispersed phase is accompanied by a significant decrease in turbulence fluctuations.} 

To quantitatively characterize the effect of dispersed phase density and viscosity in drag modulation, we have derived a conserved quantity $J^\omega$ that characterizes the radial transport of azimuthal momentum in the two-phase Taylor-Couette turbulence
 \begin{equation}
  J^\omega=J_{adv}^\omega(r) + J_{dif}^\omega(r)+J_{int}^\omega(r),
  \label{equ:momentum transport}
\end{equation}
and provided the explicit expresses for the three terms on the right-hand side of Eq.~(\ref{equ:momentum transport}), which are contributions from advection, diffusion, and two-phase interface, respectively (see Supplemental Material for the detailed derivations)
 \begin{equation}
 J_{adv}^\omega(r)=\left \langle r^3 \rho u_r \omega \right \rangle _{A,t}, 
 \vspace{-2mm}
 \end{equation}
\begin{equation}
J_{dif}^\omega(r)=-\left \langle \mu (r^3 \partial _r \omega + r\partial _\theta u_r) \right \rangle _{A,t}, 
\vspace{-2mm}
 \end{equation}
\begin{equation}
 J_{int}^\omega(r)= -\int_{r} \left \langle {r^2 f_\theta}\right \rangle _{A,t}  \, dr.\\
\end{equation}
These three terms are closely related to density, viscosity and interfacial tension, respectively.
By normalizing Eq.~(\ref{equ:momentum transport}) with the single-phase nonvortical laminar current $J_{lam}^\omega=2\mu_f r_i^2 r_o^2 \omega_i /(r_o^2 - r_i^2)$, we obtain the Nusselt number $Nu_{\omega} = J^\omega / J_{lam}^\omega$, which represents the overall transport of azimuthal momentum. Additionally, we can express the contributions of advection ($Nu_{\omega,adv}(r)$), diffusion ($Nu_{\omega,dif}(r)$), and two-phase interface ($Nu_{\omega,int}(r)$) as functions of the radial position $r$. 
It is well-established that the Nusselt number $Nu_{\omega}$ and torque $T$ in the Taylor-Couette system are related through the equation $T=2\pi L J_{lam}^{\omega} Nu_\omega$~\citep{eckhardt2007torque}. This relationship offers a convenient way to effectively decouple the effects of density, viscosity, and two-phase interface on drag reduction (see Fig.~\ref{fig:S1} in the Supplemental Material). \textcolor{black}{Note that the advection contribution comprises both an average part and a turbulent part, with the average part being negligible compared to the turbulent part (see Fig.~\ref{fig:S2} in the Supplemental Material). Therefore, similar to that in plate flows~\citep{picano2015turbulent,wang_jiang_sun_2023}, the advection contribution may be equivalent to the turbulence contribution.}

\begin{figure}
	\centering
	\includegraphics[width=0.8\linewidth]{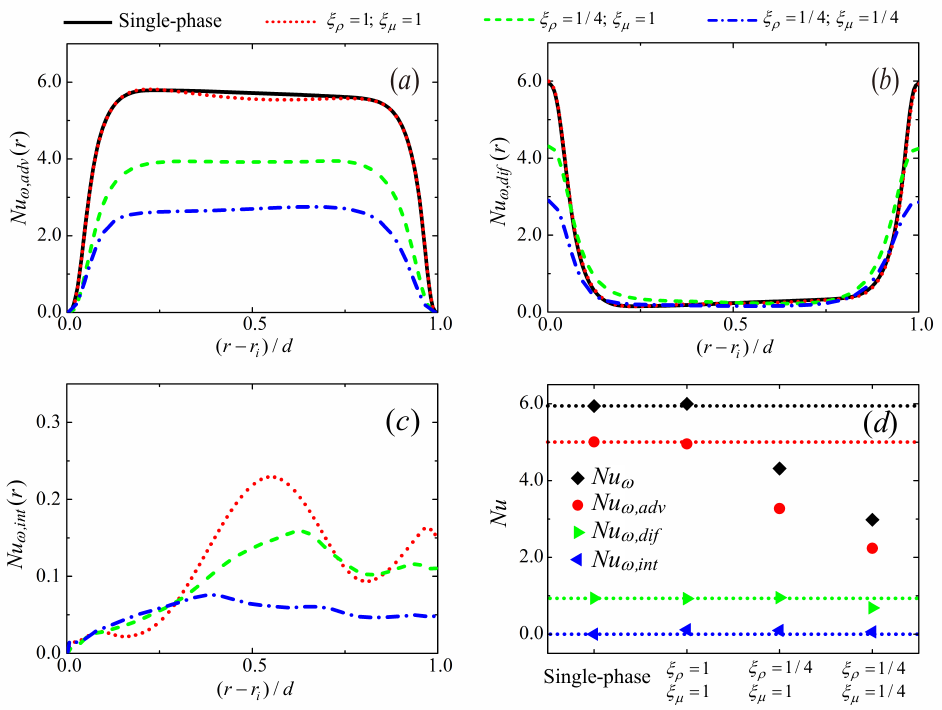}
	\caption{
		Momentum transport analysis. (\textit{a}) The normalized advection contribution, (\textit{b}) the normalized diffusion contribution, and (\textit{c}) the normalized interface contribution as a function of the radial position are shown. (\textit{d}) The momentum transport and its three contributions are averaged in the radial direction to characterize the corresponding terms within the whole system. The dashed lines in (\textit{d}) represent the averaged values for single-phase flow.
	}
	\label{fig:momentum transport}
\end{figure}

For the two-phase flow with $\xi_{\rho} = 1$ and $\xi_{\mu} = 1$, the advection contribution is slightly reduced in regions where the interface contribution is relatively large (see Fig.~\ref{fig:momentum transport}a and Fig.~\ref{fig:momentum transport}c), 
suggesting that the two-phase interface has a subtle modulation effect on the advection processes, consistent with the conservation of azimuthal momentum transport.

Considering the limited change in the diffusion contribution (see Fig.~\ref{fig:momentum transport}b), it is evident that the two-phase interface becomes the primary factor responsible for drag enhancement. 
In view of the insignificant effect of drag enhancement (see Fig.~\ref{fig:momentum transport}d), we additionally calculated a case where the Reynolds number and the surface tension coefficient are tripled, again showing the dominant role of interfacial tension in drag enhancement (see Fig.~\ref{fig:S3} in the Supplemental Material).  
\textcolor{black}{It is important to note that the interface contribution, as shown in Fig.~\ref{fig:momentum transport}c and Fig.~\ref{fig:momentum transport}d, is consistently positive, indicating that the two-phase interface does not contribute to drag reduction in the system. However, due to the small magnitude of the obtained interface contribution, which accounts for about $4\%$ of the conserved quantity, a more in-depth investigation specifically on this term should be conducted in future work.} 

Since the interface contribution ($J_{int}^\omega(r)$) acts primarily to increase drag and is very small in our drag reduction cases (see Fig.~\ref{fig:momentum transport}d), we will focus on the advection ($J_{adv}^\omega(r)$) and diffusion ($J_{dif}^\omega(r)$) contributions in the subsequent analysis.
Lowering the $\xi_{\rho}$ to 1/4 results in a reduction in the local advection contribution. The density ratio $\xi_{\rho}=1/4$ promotes dispersed phase's gathering near the inner wall, which in turn leads to a significant decrease in the upstream advection contribution. This ultimately results in an overall reduction in the advection contribution (see Fig.~\ref{fig:momentum transport}a).
Based on the conservation of momentum transport, the diffusion contribution is redistributed (see Fig.~\ref{fig:momentum transport}b) but the total diffusion contribution remains the same (see Fig.~\ref{fig:momentum transport}d). 
Further reducing $\xi_{\mu}$ to 1/4 causes a decrease in the local diffusion contribution near the inner wall (see Fig.~\ref{fig:momentum transport}b), ultimately leading to a reduction in the total diffusion contribution (see Fig.~\ref{fig:momentum transport}d). Additionally, based on the conservation of momentum transport, the advection contribution is modulated and again overall reduced (see Fig.~\ref{fig:momentum transport}a). It is evident that the combination of decreasing density ratio $\xi_{\rho}$ and viscosity ratio $\xi_{\mu}$ plays a dominant role in drag reduction. Therefore, the blocking effect of the dispersed phase on momentum transport is due to the decreased density and viscosity ratios. \textcolor{black}{In our ongoing preliminary simulations on drag modulation in two-phase plane Couette turbulence, we have observed similar drag modulation mechanisms. This is beyond the scope of the current work and will be systematically investigated and compared with the present study in our future work.}

Given the generality of Eq.~(\ref{equ:momentum transport}), our conclusion can be extended to analyze bubble drag reduction in turbulent flows~\citep{van2013importance}. Due to the extremely low density and viscosity ratios of the bubble to the continuous phase, the advection and diffusion contributions are significantly reduced, resulting in drag reduction. 
Meanwhile, it has been reported that bubbly drag reduction is effective when large bubbles are present, but the drag reduction effect is lost when the bubbles are reduced to small sizes after the addition of surfactants~\citep{verschoof2016bubble}. We can provide a physical understanding of this observation. When surfactants are added to decrease the surface tension coefficient, it leads to a reduction in bubble size. However, this reduction in bubble size is accompanied by an increase in the interfacial area in the system. As a result, the interface contribution to drag, which is enhanced by the increased interfacial area, offsets the drag reduction caused by the low density and viscosity ratios. Furthermore, the decrease in bubble size also attenuates the effective centripetal Froude number exerted on the small bubbles~\citep{van2013importance}, leading to a reduced accumulation of gas near the inner wall.
This can explain why the drag reduction is lost when the bubbles are shrunk to small sizes after the addition of surfactants.

\section {Conclusions}
         
In conclusion, we have derived a conserved quantity that characterizes the radial transport of azimuthal momentum in the fluid-fluid two-phase Taylor-Couette turbulence. This conserved quantity consists of three terms: the density-related advection contribution, the viscosity-related diffusion contribution, and the interface contribution. Our analysis highlights the significant roles played by two-phase interface, density and viscosity ratios in modulating drag.
Specifically, decreasing the density ratio of the dispersed phase to the continuous phase reduces the local advection contribution, while decreasing the viscosity ratio reduces the local diffusion contribution. Through modulation and redistribution, these effects lead to an overall reduction in momentum transport. On the other hand, the two-phase interface consistently produces a positive contribution to drag enhancement.
By considering the interplay between density ratio, viscosity ratio and two-phase interface, we conclude that drag modulation is achieved through the combined influence of these factors. The current findings contribute to a better understanding of the mechanisms underlying drag reduction in two-phase turbulent flows.  

%\backsection[Supplementary data]{\label{SupMat}Supplementary material and movies are available at \\https://doi.org/10.1017/jfm.2019...}

%\backsection[Acknowledgements]{Acknowledgements may be included at the end of the paper, before the References section or any appendices. Several anonymous individuals are thanked for contributions to these instructions.}

\backsection[Funding]{This work is financially supported by the National Natural Science Foundation of China under Grant No. 11988102 and the New Cornerstone Science Foundation through the New Cornerstone Investigator Program and the XPLORER PRIZE.}

\backsection[Declaration of interests]{ The authors report no conflict of interest.}

\backsection[Author ORCID]{\protect\\
Junwu Wang, https://orcid.org/0000-0003-3988-1477;\protect\\ 
Chao Sun, https://orcid.org/0000-0002-0930-6343.}

\section*{Supplemental Material}
\makeatletter
\renewcommand{\maketag@@@}[1]{\hbox{\m@th\normalsize\normalfont#1}}%
\makeatother
\setcounter{equation}{0}
\renewcommand{\theequation}{S\arabic{equation}}
\setcounter{figure}{0}
\renewcommand{\thefigure}{S\arabic{figure}}
\setcounter{table}{0}
\renewcommand{\thetable}{S\arabic{table}}
\subsection*{Numerical methods and computational accuracy}
\begin{figure*}
	\centering
	\includegraphics[width=0.7\linewidth]{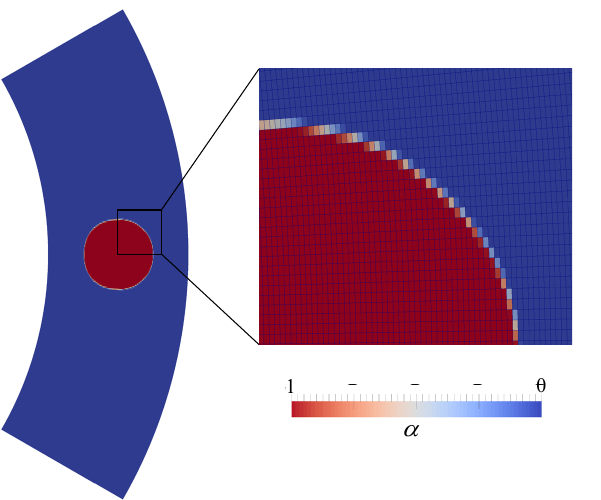}
	\caption{
		 \textcolor{black}{A snapshot of the two-phase interface resolved by the PLIC algorithm.}
	}
	\label{fig:S01}
\end{figure*}

\textcolor{black}{
The simulations are performed using a volume-of-fluid (VOF) method with a piecewise-linear interface calculation (PLIC) algorithm, which is implemented in the interFoam solver of the open-source OpenFOAM v8. In the VOF method, the phase fraction variable $\alpha$ is utilized in each cell to characterize the distribution of the two phases. The range of $\alpha$ is from zero to one, where $\alpha=0$ represents the continuous phase, $\alpha=1$ represents the dispersed phase, and $0<\alpha<1$ represents the interface region.
The evolution of $\alpha$ is governed by the transport equation
\begin{equation}
  {\partial _t \alpha}
  +\nabla\cdot(\alpha \mathbf{u})
  =0,
  \vspace{-0mm}
\end{equation}
where $\mathbf{u}$ is the velocity field. Because of the continuity of the phase fraction, the interface between the two phases tends to become smeared. To mitigate this issue, previous versions of OpenFOAM implemented an interface compression approach based on the counter-gradient transport to keep the sharpness of the interface~\citep{weller2008new}. 
In addition, the multidimensional universal limiter with explicit solution (MULES) algorithm is implemented to ensure that the phase fraction $\alpha$ remains within the strict bounds of 0 and 1.
After adding the interface compression term which is only active at the interface, the transport equation becomes
\begin{equation} 
  {\partial _t \alpha}
  +\nabla\cdot(\alpha \mathbf{u})
  +\nabla\cdot[\alpha(1-\alpha) \mathbf{u}_c]
  =0,
  \vspace{-0mm}
\end{equation}
where $\mathbf{u}_c = c\mathbf{u}\nabla{\alpha}/|\nabla{\alpha}|$ with $c$ being the compression factor.
Alternatively, a PLIC-based algorithm has been recently implemented to capture the interface more accurately. 
This algorithm involves representing the interface between the two phases by employing surface-cuts which split each cell to match the volume fraction of the phase in that cell. The surface-cuts are oriented according to the point field of the local phase fraction. The phase fraction on each cell face is then calculated from the amount submerged below the surface-cut. Note that this algorithm may cannot handle certain cells when the cut position is unclear or multiple interfaces exist. In such cases, the interface compression approach is still applied to those cells. Compared to the traditional PLIC method, the present approach is easier to implement when using an unstructured mesh. Moreover, when solving practical engineering problems, the combination of the present method with the interface compression approach enhances the robustness of the solutions. Therefore, we have employed this PLIC-based VOF method in our study to deal with the two-phase turbulence in TC system. Figure~\ref{fig:S01} shows a snapshot of a droplet placed in the Taylor-Couette system. The resolved interface region ($0<\alpha<1$) is confined within a single layer of grid cells between the two phases to ensure sharpness of the interface. }

\begin{figure*}
	\centering
	\includegraphics[width=0.95\linewidth]{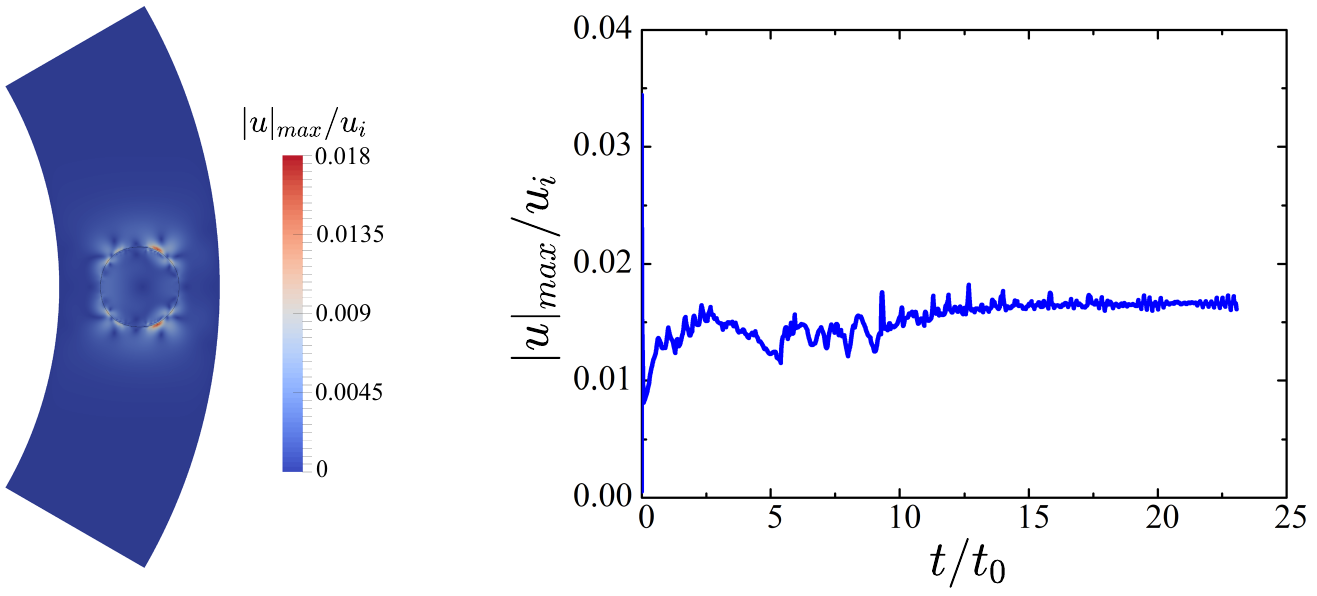}
	\caption{
		 \textcolor{black}{Spurious currents for a static drop in a Taylor-Couette system with the two cylinders fixed. (\textit{a}) Contour of the velocity magnitude. (\textit{b}) The maximum velocity magnitude as a function of time. The maximum velocity magnitude is normalized by the velocity of the inner cylinder $u_i$ considered in our work and the time t is normalized by the time required for one turn of the inner cylinder $t_0=2\pi r_i/u_i$, where $r_i$ is the radius of the inner cylinder.}
	}
	\label{fig:S02}
\end{figure*}

\textcolor{black}{
Spurious currents have long been a significant obstacle in relevant research, affected by factors including surface tension coefficient, grid resolution, time step, density ratio, and viscosity ratio between phases ~\citep{vachaparambil2019comparison,harvie2006analysis,deshpande2012evaluating}. To date, completely eliminating spurious currents remains a formidable challenge. In our study, we address this issue by carefully selecting suitable simulation conditions and parameters to ensure that spurious currents remain within an acceptable range. Figure~\ref{fig:S02} shows a snapshot of a static drop with a diameter half of the gap width. The density ratio of the drop to the continuous phase is set as 1/4 and the viscosity ratio is set as 1/4 which has strongest spurious currents considered in our work. The inner and outer cylinders are fixed and the spurious current mainly appear near the two-phase interface. The normalized maximum velocity magnitude is limited to below $1.8\%$ as shown in Fig.~\ref{fig:S02}. By doing this, the almost constant conserved quantity for two-phase flow could be obtained as shown by the black solid line in Fig.~\ref{fig:S1}. The ratio of the fluctuation and mean value of the conserved quantity is below $1.2\%$.}

\textcolor{black}{
The schemes of the time integration and spatial discretization are listed below. We utilize a blended scheme for the temporal term discretization, which lies between the first-order Euler scheme and the second-order Crank-Nicolson scheme. To ensure robustness and accuracy, we set the blending factor to 0.9. For spatial discretization, we employ a second-order linear-upwind scheme to discretize the advection term in the momentum equation. The phase fraction transport equation is solved using a piecewise-linear interface calculation (PLIC) scheme. The pressure-velocity coupling is handled using the PIMPLE algorithm. The pressure equation is solved using the Geometric Algebraic Multigrid (GAMG) solver coupled with the Simplified Diagonal-based Incomplete Cholesky (DIC) smoother. For solving velocity and phase fraction, we use an iterative solver with a symmetric Gauss-Seidel smoother. In the simulation, we maintain a tolerance of $10^{-6}$ for all variables to control the residuals, except for the phase fraction, which has a tolerance of $10^{-8}$.}

\textcolor{black}{
We have before performed simulations using OpenFOAM for single-phase flows with a range of Taylor number from $Ta=5.84\times10^3$ to $Ta=2.39\times10^7$ with the temporal term discretized using a second-order implicit backward inferencing scheme~\citep{xu2022direct,xu2023direct} and validated our results through comparisons with those from Ostilla et al.~\citep{ostilla2013optimal}. 
Considering our temporal term is discretized with a blended scheme, we have additionally simulated two cases with the Taylor number being $Ta=3.90\times10^6$ (Re = 1600) and $Ta=9.52\times10^6$ (Re = 2500) and compared our results with those from Ostilla et al~\citep{ostilla2013optimal}, as shown in Table S1. 
In our work, the minimum flow geometry with a rotational symmetry of six ($n_{sym}=6$, i.e., the azimuthal angle of the simulated domain is $\pi/3$) and an aspect ratio of $\Gamma=L/d=2\pi/3$ is selected to reduce the computational cost while not affecting the results, which has been verified by previous studies ~\citep{brauckmann2013direct,ostilla2015effects}. The $L$ denotes the axial length and $d$ denotes the gap width.
Regarding the deviation analysis, we represent the Nusselt number $Nu_\omega$ from Ostilla et al.~\citep{ostilla2013optimal} 
and OpenFOAM as $Nu_{\omega,Ot}$ and $Nu_{\omega,Op}$, respectively. 
The deviation is obtained by $Nu_{\omega,Op}/Nu_{\omega,Ot}-1$ at the same Reynolds number. A deviation of $+1.13\%$ is found at Re=1600 and a deviation of $-1.37\%$ is found at Re = 2500, indicating that the chosen time scheme is sufficient to capture the flow field information.}

\begin{table}%The best place to locate the table environment is directly after its first reference in text
    	\centering
     \setlength\tabcolsep{8pt}
		\begin{tabular}{cccccccc}
			\textrm{}&
			\textrm{$Re$}&
                \textrm{$Ta$}&
                \textrm{$n_{sym}$}&
                 \textrm{$\Gamma$}&
                 \textrm{$N_\theta \times N_r \times N_z$}&
                 \textrm{$Nu_\omega$}&
			\multicolumn{1}{c}{Deviation}\\[6pt]
			%\colrule
			%\hline
			Ostilla et al.    & 1600	& $3.90\times10^6$ & $1$	    & $2\pi$		 & $300\times144\times144$	    & 5.42553     & --- \\[2pt]
			OpenFOAM          & 1600 	& $3.90\times10^6$ & $6$	    & $2\pi/3$       & $80\times160\times80$	    & 5.48683     & $+1.13\%$\\[2pt]
			Ostilla et al.    & 2500 	& $9.52\times10^6$ & $1$	    & $2\pi$         & $384\times192\times192$	    & 6.42160     & --- \\[2pt]
			OpenFOAM          & 2500 	& $9.52\times10^6$ & $6$ 	& $2\pi/3$       & $100\times192\times100$	    & 6.33191     & $-1.37\%$ \\[2pt]
		\end{tabular}
     	\caption{\label{tab:table S1}%
		\textcolor{black}{Validity of the calculation results for the single-phase flow at Re = 1600 and Re = 2500.}
	}
\end{table}

\subsection*{Formula derivation of the constant azimuthal momentum transport}
The azimuthal momentum in a Taylor-Couette flow in the cylindrical coordinate system ($r$, $\theta$, $z$) is governed by
\begin{small} 
\begin{equation}
\begin{split}
\displaystyle\frac{\partial(\rho u_\theta)}{\partial t}
+\displaystyle\frac{2\rho u_r u_\theta}{r}
+\displaystyle\frac{\partial (\rho u_r u_\theta)}{\partial r}
+\displaystyle\frac{1}{r} \displaystyle\frac{\partial (\rho u_\theta u_\theta)}{\partial \theta} 
+\displaystyle\frac{\partial (\rho u_\theta u_z)}{\partial z}
=  
\displaystyle\frac{2}{r}
\Big(
\displaystyle\frac{\mu}{r}\displaystyle\frac{\partial u_r}{\partial \theta}
+\displaystyle\frac{\mu \partial u_\theta}{\partial r}
-\displaystyle\frac{\mu u_\theta}{r}
\Big) 
+\\\displaystyle\frac{\partial}{\partial r}
\Big(
\displaystyle\frac{\mu}{r}\displaystyle\frac{\partial u_r}{\partial \theta}
+\displaystyle\frac{\mu \partial u_\theta}{\partial r}
-\displaystyle\frac{\mu u_\theta}{r}
\Big)
+\displaystyle\frac{1}{r}\displaystyle\frac{\partial}{\partial \theta}
\Big(
\displaystyle\frac{2\mu}{r}\displaystyle\frac{\partial u_\theta}{\partial \theta}
+\displaystyle\frac{2\mu u_r}{r}
\Big)
+\displaystyle\frac{\partial}{\partial z}
\Big(
\displaystyle\frac{\mu}{r}\displaystyle\frac{\partial u_z}{\partial \theta}
+\displaystyle\frac{\mu \partial u_\theta}{\partial z}
\Big)
-\displaystyle\frac{1}{r}\displaystyle\frac{\partial p}{\partial \theta}
+f_\theta,
\end{split}
\label{equ:S1}
\end{equation}
\end{small}%%%
where $u_r$, $u_\theta$, and $u_z$ are the radial velocity, the azimuthal velocity and the axial velocity, respectively. $f_\theta$ is the azimuthal interfacial tension. Although the density and viscosity of the dispersed phase ($\rho_d$ and $\mu_d$) and carrier phase ($\rho_f$ and $\mu_f$) are constants, the effective density  $\rho$ and the effective viscosity  $\mu$ are variables in two-phase flow and are defined as $\rho=\alpha\rho_d+(1-\alpha)\rho_f $ and $\mu=\alpha\mu_d+(1-\alpha)\mu_f $ in the volume-of-fluid method, where $\alpha$ is the phase fraction of dispersed phase. We apply the following operator to Eq.~(\ref{equ:S1}):
\begin{small} 
\begin{equation}
\left \langle \cdot \right \rangle _{A, t}
=
\displaystyle\frac{1}{2\pi L T}
\int_{0}^{T}\,
\int_{0}^{2\pi}\,
\int_{0}^{L}\,
dz d\theta dt,
\label{equ:S2}
\end{equation}
\end{small}%% 
i.e., we average all the quantities over time, axially, and azimuthally. $T$ is the total time and $L$ is the height of Taylor-Couette system. Given the axial periodicity, azimuthal periodicity, and statistical steady state, any term in Eq.~(\ref{equ:S1}) of the form $\displaystyle\frac{\partial X}{\partial z}$, $\displaystyle\frac{\partial X}{\partial \theta}$, and $\displaystyle\frac{\partial X}{\partial t}$ will be 0 once integrated. The Eq.~(\ref{equ:S1}) is therefore rewritten as
\begin{small} 
\begin{equation}
\begin{split}
\left \langle 
\displaystyle\frac{2\rho u_r u_\theta}{r}
+\displaystyle\frac{\partial (\rho u_r u_\theta)}{\partial r}
\right \rangle _{A, t}
=  
\left \langle 
\displaystyle\frac{2}{r}
\Big(
\displaystyle\frac{\mu}{r}\displaystyle\frac{\partial u_r}{\partial \theta}
+\displaystyle\frac{\mu \partial u_\theta}{\partial r}
-\displaystyle\frac{\mu u_\theta}{r}
\Big) 
+\displaystyle\frac{\partial}{\partial r}
\Big(
\displaystyle\frac{\mu}{r}\displaystyle\frac{\partial u_r}{\partial \theta}
+\displaystyle\frac{\mu \partial u_\theta}{\partial r}
-\displaystyle\frac{\mu u_\theta}{r}
\Big)
+f_\theta
\right \rangle _{A, t}.
\label{equ:S3}
\end{split}
\end{equation}
\end{small}%% 
Combining the two terms on the left side of Eq.~(\ref{equ:S3}), we have
\begin{small} 
\begin{equation}
\begin{split}
\left \langle 
\displaystyle\frac{2\rho u_r u_\theta}{r}
+\displaystyle\frac{\partial (\rho u_r u_\theta)}{\partial r}
\right \rangle _{A, t}
=
\left \langle 
\displaystyle\frac{1}{r^2}
\bigg(
2r\rho u_r u_\theta
+\displaystyle\frac{r^2\partial (\rho u_r u_\theta)}{\partial r}
\bigg)
\right \rangle _{A, t}
=
\left \langle 
\displaystyle\frac{1}{r^2}\displaystyle\frac{\partial (r^2\rho u_r u_\theta)}{\partial r}
\right \rangle _{A, t}.
\label{equ:S4}
\end{split}
\end{equation}
\end{small}%% 
The terms on the right side of Eq.~(\ref{equ:S3}) can be rewritten as 
\begin{small} 
\begin{equation}
\begin{split}
\left \langle 
\displaystyle\frac{2\mu}{r^2}\displaystyle\frac{\partial u_r}{\partial \theta}
+\displaystyle\frac{2\mu}{r}\displaystyle\frac{\partial u_\theta}{\partial r}
-\displaystyle\frac{2\mu u_\theta}{r^2}
-\displaystyle\frac{\mu}{r^2}\displaystyle\frac{\partial u_r}{\partial \theta}
+\displaystyle\frac{\mu}{r}\displaystyle\frac{\partial u_r}{\partial r \partial \theta} 
 +\displaystyle\frac{\mu \partial^2 u_\theta}{\partial r^2}
 \right \rangle _{A, t} \\
+\left \langle 
\displaystyle\frac{\mu u_\theta}{r^2}
-\displaystyle\frac{\mu}{r}\displaystyle\frac{\partial u_\theta}{\partial r}
+\displaystyle\frac{1}{r}\displaystyle\frac{\partial \mu}{\partial r}\displaystyle\frac{\partial u_r}{\partial \theta}
+\displaystyle\frac{\partial \mu}{\partial r}\displaystyle\frac{\partial u_\theta}{\partial r}
-\displaystyle\frac{\partial \mu}{\partial r}\displaystyle\frac{u_\theta}{r}
+f_\theta
\right \rangle _{A, t}.
\label{equ:S5}
\end{split}
\end{equation}
\end{small}%% 
Combining the similar items, including the first and fourth items, the second and eighth items, and the third and seventh items, the Eq.~(\ref{equ:S5}) can be rewritten as
\begin{small} 
\begin{equation}
\begin{split}
\left \langle 
\displaystyle\frac{\mu}{r^2}\displaystyle\frac{\partial u_r}{\partial \theta}
+\displaystyle\frac{\mu}{r}\displaystyle\frac{\partial u_\theta}{\partial r}
-\displaystyle\frac{\mu u_\theta}{r^2}
+\displaystyle\frac{\mu}{r}\displaystyle\frac{\partial u_r}{\partial r \partial \theta}
+\displaystyle\frac{\mu \partial^2 u_\theta}{\partial r^2}
+\displaystyle\frac{1}{r}\displaystyle\frac{\partial \mu}{\partial r}\displaystyle\frac{\partial u_r}{\partial \theta}
+\displaystyle\frac{\partial \mu}{\partial r}\displaystyle\frac{\partial u_\theta}{\partial r}
-\displaystyle\frac{\partial \mu}{\partial r}\displaystyle\frac{u_\theta}{r}
+f_\theta
\right \rangle _{A, t}.
\label{equ:S6}
\end{split}
\end{equation}
\end{small}%% 
Combining the second, third, fifth, seventh, and eighth items of Eq.~(\ref{equ:S6}), we have
\begin{small} 
\begin{equation}
\begin{split}
\left \langle 
\displaystyle\frac{\mu}{r}\displaystyle\frac{\partial u_\theta}{\partial r}
-\displaystyle\frac{\mu u_\theta}{r^2}
+\displaystyle\frac{\mu \partial^2 u_\theta}{\partial r^2}
+\displaystyle\frac{\partial \mu}{\partial r}\displaystyle\frac{\partial u_\theta}{\partial r}
-\displaystyle\frac{\partial \mu}{\partial r}\displaystyle\frac{u_\theta}{r}
\right \rangle _{A, t}
=
\left \langle 
\displaystyle\frac{1}{r^2}\displaystyle\frac{\partial}{\partial r}
\bigg(
\mu r^3 \displaystyle\frac{\partial }{\partial r}
\Big(
\displaystyle\frac{u_\theta}{r}
\Big)
\bigg)
\right \rangle _{A, t}.
\label{equ:S7}
\end{split}
\end{equation}
\end{small}%% 
Combining the remaining items of Eq.~(\ref{equ:S6}), we have
\begin{small} 
\begin{equation}
\begin{split}
\left \langle
\displaystyle\frac{\mu}{r^2}\displaystyle\frac{\partial u_r}{\partial \theta}
+\displaystyle\frac{\mu}{r}\displaystyle\frac{\partial u_r}{\partial r \partial \theta}
+\displaystyle\frac{1}{r}\displaystyle\frac{\partial \mu}{\partial r}\displaystyle\frac{\partial u_r}{\partial \theta}
+f_\theta
\right \rangle _{A, t}
=
\left \langle 
\displaystyle\frac{1}{r^2}\displaystyle\frac{\partial}{\partial r}
\Big(
\mu r \displaystyle\frac{\partial u_r}{\partial \theta}
\Big)
+f_\theta
\right \rangle _{A, t}.
\label{equ:S8}
\end{split}
\end{equation}
\end{small}%% 
We here introduce the operator ${\partial_x}=\displaystyle\frac{\partial}{\partial x}$ and the angular velocity $\omega=\displaystyle\frac{u_\theta}{r}$. Combining Eq.~(\ref{equ:S4}, \ref{equ:S7}, \ref{equ:S8}) and rearranging, we have
\begin{small} 
\begin{equation}
\begin{split}
\left \langle 
\displaystyle\frac{1}{r^2}{\partial_r (r^3\rho u_r \omega)}
\right \rangle _{A, t}
=
\left \langle 
\displaystyle\frac{1}{r^2}{\partial_r}(\mu r^3 {\partial_r \omega})
\right \rangle _{A, t}
+
\left \langle 
\displaystyle\frac{1}{r^2}{\partial_r}(\mu r {\partial_\theta u_r})
+f_\theta
\right \rangle _{A, t},
\label{equ:S9}
\end{split}
\end{equation}
\end{small}%% 
\begin{small} 
\begin{equation}
\begin{split}
\left \langle 
{\partial_r (r^3\rho u_r \omega)}
\right \rangle _{A, t}
=
\left \langle 
{\partial_r}(\mu r^3 {\partial_r \omega})
\right \rangle _{A, t}
+
\left \langle 
{\partial_r}(\mu r {\partial_\theta u_r})
+r^2 f_\theta
\right \rangle _{A, t}.
\label{equ:S10}
\end{split}
\end{equation}
\end{small}%% 
The Eq.~(\ref{equ:S10}) can be rewritten as
\begin{small} 
\begin{equation}
\begin{split}
\partial_r 
\left \langle r^3\rho u_r \omega \right \rangle _{A, t}
- \left \langle\mu (r^3 {\partial_r \omega}
+ r {\partial_\theta u_r}) \right \rangle _{A, t}
-\int_{r} \left \langle {r^2 f_\theta} \right \rangle _{A, t} \, dr
=0,
\label{equ:S11}
\end{split}
\end{equation}
\end{small}%%
where the  $\int_{r} \left \langle {r^2 f_\theta} \right \rangle _{A, t} \, dr$ is the integral of $ \left \langle {r^2 f_\theta} \right \rangle _{A, t}$ in the radial direction. We here define $J_{adv}^\omega(r)=\left \langle r^3 \rho u_r \omega \right \rangle _{A,t}$,  $J_{dif}^\omega(r)=-\left \langle \mu (r^3 \partial _r \omega + r\partial _\theta u_r) \right \rangle _{A,t}$, and $J_{int}^\omega(r)= -\int_{r} \left \langle {r^2 f_\theta}\right \rangle _{A,t}  \, dr$, respectively.
We can now integrate the Eq.~(\ref{equ:S11}) from the inner cylinder $r=r_i$ to a generic cylindrical surface $r=r^*$
\begin{small} 
\begin{equation}
\begin{split}
J_{adv}^\omega(r_i)
+J_{dif}^\omega(r_i)
+J_{int}^\omega(r_i)
=
J_{adv}^\omega(r^*)
+J_{dif}^\omega(r^*)
+J_{int}^\omega(r^*).
\label{equ:S12}
\end{split}
\end{equation}
\end{small}%% 
$J_{adv}^\omega(r_i)=0$ due to ${u_r}=0$ at the inner cylinder, and $J_{int}^\omega(r_i)=0$ due to ${f_\theta}=0$ at the inner cylinder. We can therefore introduce the constant quantity $J^\omega=J_{dif}^\omega(r_i)$ to characterize the transverse current of
azimuthal momentum, i.e.,
\begin{small} 
\begin{equation}
\begin{split}
J^\omega
=
J_{adv}^\omega(r^*)
+J_{dif}^\omega(r^*)
+J_{int}^\omega(r^*).
\label{equ:S13}
\end{split}
\end{equation}
\end{small}%% 
The three terms on the right side of Eq.~(\ref{equ:S13}) are the density-related advection contribution, the viscosity-related diffusion contribution, and the interface contribution, respectively. The interface contribution exhibits a distinct form from other two terms, and it is necessary to discuss its physical meaning. For the region between the arbitrary cylindrical surface and the inner cylinder, the interface acts as a source term for momentum transfer. This means that the interface contribution comes from all the interfaces in the region between the cylindrical surface and the inner cylinder. Therefore, it is necessary to sum the influence of all interfaces in the region. 

\begin{figure*}
	\centering
	\includegraphics[width=0.5\linewidth]{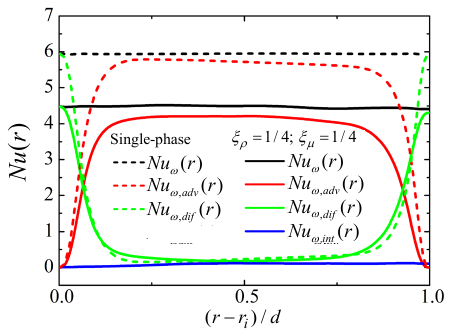}
	\caption{ {Momentum transport analysis at ${\rm Re} = 2000$. The dotted line denotes the single-phase flow and the solid line denotes the two-phase flow with $\varphi = 10\%$. }
	}
	\label{fig:S1}
\end{figure*}

The Eq.~(\ref{equ:S13}) is normalized by the single-phase laminar current $J_{lam}^\omega=2\mu_f r_i^2 r_o^2 \omega_i /(r_o^2 - r_i^2) $ to obtain the constant Nusselt number $Nu_{\omega} = J^\omega / J_{lam}^\omega$ and its three contributions related to $r$ including $Nu_{\omega,adv}(r)$ , $Nu_{\omega,dif}(r)$, and $Nu_{\omega,int}(r)$, where $r_o$ and $\omega_i$ are the radius of outer cylinder and the angular velocity of inner cylinder, respectively. The $Nu_{\omega}$ and $T$ are related by $T=2{\pi}L J_{lam}^{\omega} Nu_\omega$, providing the opportunity to effectively decouple the effects of density, viscosity, and interface structure on the drag modulation. Figure~\ref{fig:S1} shows the $Nu_{\omega}$ and its three contributions as a function of the radial position for two-phase flow with $\rho_{d}=\rho_{f}/4$ and $\mu_{d}=\mu_{f}/4$, and compares them to those for single-phase flow. Given that the value of $Nu_{\omega}$ hardly changes with $r$, it can be sure that present simulations accurately capture the momentum transfer process.  The dispersed phase shows to alter the density-related advection contribution and the viscosity-related diffusion contribution, and induce non-zero interface contribution. By varying density or viscosity of the dispersed phase individually, it becomes possible to reveal their respective roles in drag modulation.
%\vspace*{40\baselineskip} 
\begin{figure*}
	\centering
	\includegraphics[width=1\linewidth]{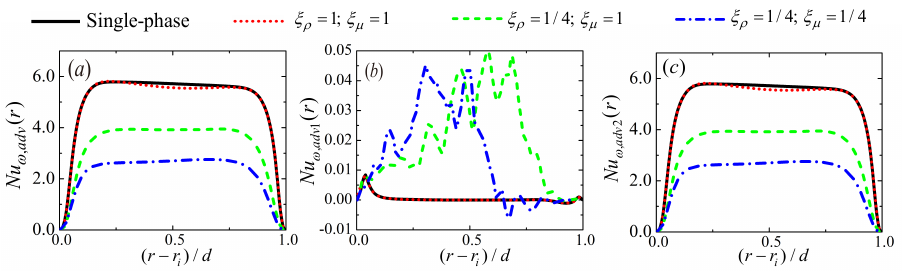}
	\caption{
		 \textcolor{black}{Advection contribution analysis. (\textit{a}) The whole advection contribution, (\textit{b}) the average part, and (\textit{c}) the turbulent part are shown as a function of the radial position.}
	}
	\label{fig:S2}
\end{figure*} 
\begin{figure*}
	\centering
	\includegraphics[width=0.6\linewidth]{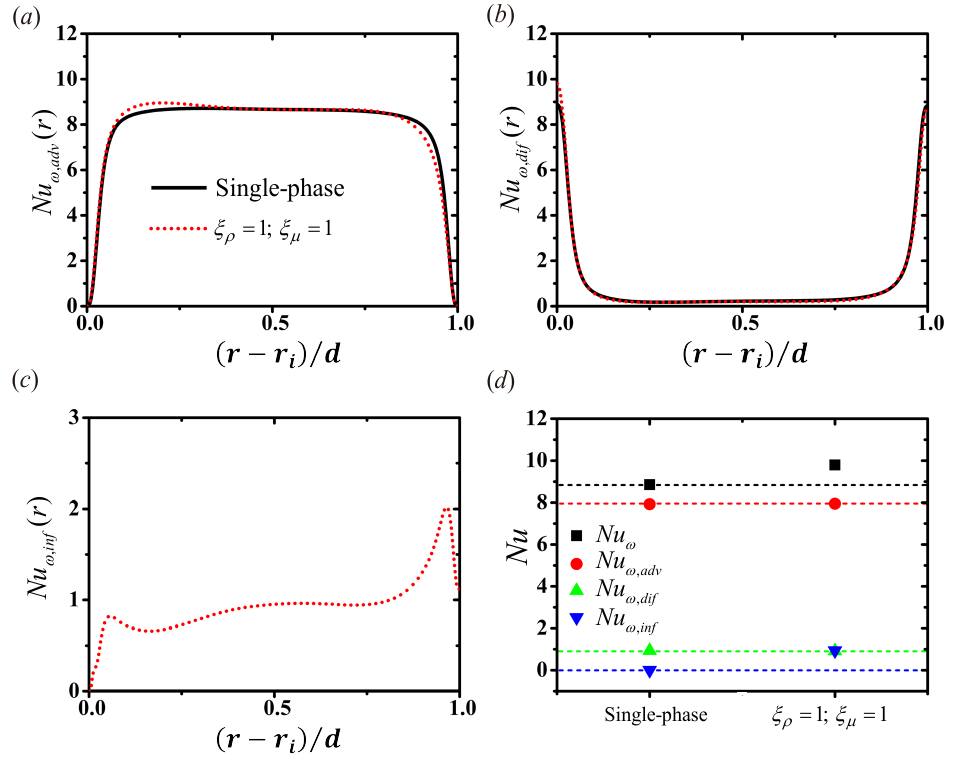}
	\caption{
		Momentum transport analysis for ${\rm Re}=6000$ and $We=3915$. (\textit{a}) The normalized advection contribution, (\textit{b}) the normalized diffusion contribution, and (\textit{c}) the normalized interface contribution as a function of the radial position are shown. (\textit{d}) The momentum transport and its three contributions are averaged in the radial direction. The dashed lines in (\textit{d}) represent the averaged values for single-phase flow.
	}
	\label{fig:S3}
\end{figure*} 

\textcolor{black}{Strictly speaking, based on the Favre averaging, the formula of the constant quantity can be rewritten as
\begin{small} 
\begin{equation}
\begin{split}
J^\omega
=
J_{adv1}^\omega(r)
+J_{adv2}^\omega(r)
+J_{dif}^\omega(r)
+J_{int}^\omega(r),
\label{equ:S14}
\end{split}
\end{equation}
\end{small}%% 
where 
$J_{adv1}^\omega(r)
=
{r^2 \overline{\rho} \widetilde{u_r} \widetilde{u_\theta}}$ and
$J_{adv2}^\omega(r)
=
{\left \langle {r^2\rho u_r^{\prime\prime} u_\theta^{\prime\prime}} \right \rangle _{A, t}}$ 
are the average part and the turbulent part of the advection term $J_{adv1}^\omega(r)$, respectively. Here,
$\overline{\rho}
=
\left \langle {\rho} \right \rangle _{A,t}$, 
$\widetilde{u_\theta}
=
\left \langle {\rho u_\theta} \right \rangle _{A,t}
/
\left \langle {\rho} \right \rangle _{A,t}$ 
and $\widetilde{u_r}
=
\left \langle {\rho u_r} \right \rangle _{A,t}
/
\left \langle {\rho} \right \rangle _{A,t}$.The two parts of the advection contribution are normalized to obtain  $Nu_{\omega,adv1}(r) = J_{adv1}^\omega(r) / J_{lam}^\omega$ and $Nu_{\omega,adv2}(r) = J_{adv2}^\omega(r) / J_{lam}^\omega$, which are shown in Fig.~\ref{fig:S2}. The average part of the advection contribution is negligible compared to the turbulent part since their maximum ratio is within $1.7\%$ and the mean value of the ratio is about $0.59\%$. Hence, the advection contribution, the primary focus of our manuscript, is synonymous with the turbulence contribution.} 

%\backsection[Author contributions]{Authors may include details of the contributions made by each author to the manuscript, for example, ``A.G. and T.F. derived the theory and T.F. and T.D. performed the simulations.  All authors contributed equally to analysing data and reaching conclusions, and in writing the paper.''}

\bibliographystyle{jfm}
\bibliography{jfm}

%Use of the above commands will create a bibliography using the .bib file. Shown below is a bibliography built from individual items.

%\bibliographystyle{jfm}
%\bibliography{jfm2esam}

%% End of file `jfm2esam.bib'.

\end{document}